\documentstyle[aps,preprint]{revtex}
\begin{document}

\draft

\title{Nonlinear vertical oscillations of a particle in a sheath
\\of a rf discharge}

\author{A. V. Ivlev,\footnotemark[1] R. S\"utterlin,
V. Steinberg,\footnotemark[2] M. Zuzic and G. Morfill }

\address{Max-Planck-Institut f\"ur Extraterrestrische Physik,
D-85740 Garching, Germany}

\footnotetext[1]{Permanent address: High Energy Density Research
Center RAS, 127412 Moscow, Russia.} \footnotetext[2]{Present
address: Department of Physics of Complex Systems, Weizmann
Institute of Science, Rehovot 76100, Israel.}

\maketitle

\begin{abstract}
A new simple method to measure the spatial distribution of the
electric field in the plasma sheath is proposed. The method is
based on the experimental investigation of vertical oscillations
of a single particle in the sheath of a low-pressure
radio-frequency discharge. It is shown that the oscillations
become strongly nonlinear and secondary harmonics are generated as
the amplitude increases. The theory of anharmonic oscillations
provides a good qualitative description of the data and gives
estimates for the first two anharmonic terms in an expansion of
the sheath potential around the particle equilibrium.
\end{abstract}


There are various experimental methods on the plasma sheath
diagnostic in low-pressure gas discharges (see, e.g.,
\cite{Czar,Sato,Mahony} and references therein). Most of the
methods are based on optical emission spectroscopy and temporally
resolved probe measurements and are rather complicated
technically. We propose an ``alternative'' simple method to
measure the spatial distribution of the electric field in the
plasma sheath based on investigation of ``large-amplitude''
vertical oscillations of micron sized particles.

So far, investigation of the particle oscillations in the sheath
of a radio-frequency (rf) discharge
\cite{Nitter2,Zuzic,Nunomura,Ivlev} was one of the methods used to
determine the particle charge \cite{Melzer,Trott,Scholl}. This
method is based on the assumption that the vertical distribution
of the sheath potential {\it is known} and can be approximated by
a parabolic profile \cite{Beatrice}, so that the oscillations are
always harmonic. This assumption is reasonable for sufficiently
high gas pressure, when the mean free path of ions is much less
than the vertical spatial scale of the sheath field
\cite{Beatrice}: For example, the model of a parabolic rf sheath
in an argon plasma can be applied for pressures above
$\simeq20$~Pa. But for $p\lesssim 10$~Pa the sheath profile might
deviate strongly from a parabolic one. In this case, the
oscillations are harmonic only if their amplitude is very small.

In this Letter, we report experimental and theoretical
investigations of nonlinear (large-amplitude) vertical
oscillations of a single particle in the sheath of a low-pressure
rf discharge. We compare the experimental results with the theory
of anharmonic oscillations and estimate the first two anharmonic
terms in an expansion of the sheath potential around the particle
equilibrium height. Thus, the study of large-amplitude particle
oscillations allows us to reconstruct the distribution of the
vertical electric field in wide region of the sheath.

{\it Experimental setup and results} --- The experiments are
performed in a so-called Gaseous Electronic Conference (GEC) rf
reference cell \cite{GEC} with the electrode system modified as
shown in Fig.~\ref{f1}.  The lower aluminum electrode, 230 mm in
diameter, is capacitively coupled to a rf generator providing a
peak-to-peak voltage $\simeq70$~V (rf power $\simeq2$ W). A
ring-shaped upper grounded electrode with an outer diameter of
200~mm and a thickness of 10~mm is located 50~mm above the lower
electrode. A 2~mm thick copper ring with a 40~mm inner diameter is
placed on the lower electrode to confine the particle above the
center. Argon gas at a pressure 0.5~Pa is used for the discharge.
For this pressure, a self-bias voltage $\simeq-33$~V on the lower
electrode provides the sheath with a (visible) thickness of
$\simeq20$~mm and causes a spherical polystyrene particle (density
$1.05$~g/cm$^3$, diameter $7.6\pm 0.1~\mu$m) to levitate
$\simeq14$~mm above the electrode.

At a height 8~mm above the lower electrode a horizontal wire of
0.5~mm diameter and 80~mm length is placed directly below the
levitated particle. A low-frequency sinusoidal voltage of
amplitude $U_{\rm ex}$ is applied to this wire to excite the
vertical oscillations of the particle. The frequency $\omega/2\pi$
of the voltage could be varied from 0.1~Hz up to 40~Hz and back in
0.1~Hz steps. To determine the oscillation amplitude at each step,
we illuminated the particle with a vertical laser sheet of
$\simeq140~\mu$m thickness and imaged it from the side by an
external digital CCD camera (max. field rate is 50 fields/s,
resolution is $768\times576$ pixels).

For relatively small values of the excitation voltage ($U_{\rm
ex}\lesssim30$~mV), we observed the usual linear (harmonic)
oscillations: The dependence of the oscillation amplitude on the
frequency has a single symmetric narrow maximum at the resonance
frequency $\omega_0/2\pi\simeq 17.0$~Hz (primary resonance).
However, for larger $U_{\rm ex}$ the dependence clearly exhibits
nonlinearity of the oscillations (see, e.g., \cite{Land}).
Figure~\ref{f2} shows that the maximum of the primary resonance
shifts to the left, as the excitation voltage increases. When
$U_{\rm ex}$ exceeds some critical value, a hysteresis appears. To
illustrate this phenomenon, we can compare Figs.~\ref{f2}(a) and
\ref{f2}(b) ($U_{\rm ex}=100$~mV and 200~mV). When the frequency
increases [Fig.~\ref{f2}(a)], the amplitude grows continuously
until a certain frequency ($\omega_+$) is reached, then it
``jumps'' upward and thereafter decreases monotonically as
$\omega$ is further increased. On the ``way back''
[Fig.~\ref{f2}(b)], the amplitude grows continuously until another
frequency ($\omega_-$) is reached. Then it spontaneously ``jumps''
downward and continuous to decrease slowly as the frequency is
reduced further. We found that $\omega_+$ is always larger, than
$\omega_-$. Outside this ``hysteresis zone'', the experimental
points obtained for increasing and decreasing frequencies
practically overlap. The shift of the maximum, as well as the
width of the ``hysteresis zone'' rapidly increase with $U_{\rm
ex}$.

Another peculiarity of the oscillations are the secondary
resonances -- two maxima at $\omega\simeq \frac12\omega_0$ and
$\omega\simeq 2\omega_0$, which are observed for sufficiently high
excitation voltage. Figure~\ref{f3} shows the peak in the vicinity
$\omega\simeq \frac12\omega_0$ (superharmonic resonance) for
different values of $U_{\rm ex}$. We see that the dependence has a
finite offset with some slope (marked by the dashed line), which
represents the low-frequency tail of the primary resonance. The
shape of the peaks resembles that of the primary resonance: As
$U_{\rm ex}$ increases, the maximum shifts to lower $\omega$ and
the hysteresis appears. In contrast, for the subharmonic resonance
at $\omega\simeq2\omega_0$ (Fig.~\ref{f4}) the dependence is
qualitatively different. When $U_{\rm ex}$ is below a certain
threshold, we see only the offset -- the high-frequency tail of
the primary resonance -- but when $U_{\rm ex}$ exceeds this
threshold, the amplitude increases abruptly. When $\omega$ is
increased [Fig.~\ref{f4}(a)], the amplitude decreases very slowly
along the offset line up to some point. Then the amplitude jumps
upward and subsequently decreases rapidly until the resonance
branch crosses the offset line. At higher frequencies the
dependence again follows the offset. When $\omega$ is decreased
[Fig.~\ref{f4}(b)], the amplitude varies approximately as in the
case of increasing frequency, but there is no readjustment
downward anywhere. The amplitude increases until the particle is
ejected from the sheath. Both the height and the width of the peak
rapidly increase with $U_{\rm ex}$.

It is worth noting that the amplitude of oscillations both for
primary and secondary resonances is always much smaller than the
sheath thickness ($\simeq20$~mm).

{\it Theory and discussion} --- The particle oscillations are
determined by the spatial distribution of the electrostatic
potential, $\phi(z)$, in the sheath. In the absence of excitation,
a particle is levitated in the minimum of the potential well (at
the height $z=0$), where the gravitational force, $Mg$, is
balanced by the electrostatic force, $Q(0)E(0)\equiv Q_0E_0$. We
assume that the equilibrium particle charge, $Q(z)$, changes
weakly with $z$ [compared to the electric field of the sheath,
$E(z)$], so that we can set $Q\simeq Q_0=$~const. The particle
collision experiments \cite{Uwe} in Ar discharge at pressures
$p\sim1$~Pa confirm this assumption: Typically, the spatial scale
of the field change is approximately one order smaller than that
of the charge change \cite{Uwe_Diss}. The electrostatic energy of
the particle, ${\cal U}=Q\phi$, can be expanded around $z=0$ in
the series,
\begin{equation}\label{11}
{\textstyle{\cal U}(z)={\cal U}_0^{'}z+\frac12{\cal U}_0^{''}z^2+
\frac16{\cal U}_0^{'''}z^3+\frac1{24}{\cal U}_0^{(4)}z^4+O(z^5)}.
\end{equation}
Using the condition of equilibrium $Mg=QE_0\equiv-{\cal U}_0^{'}$
and the definition of the resonance frequency, $M\omega_0^2=-Q
E_0^{'}\equiv{\cal U}_0^{''}$, we rewrite it as follows:
\begin{equation}\label{1}
{\textstyle{\cal U}(z)\simeq M\left(-gz+\frac12\omega_0^2z^2+
\frac13\alpha z^3+\frac14\beta z^4\right)},
\end{equation}
where $\alpha={\cal U}_0^{'''}/2M$ and $\beta={\cal U}_0^{(4)}/
6M$ are the anharmonic coefficients. We keep only the first two
anharmonic coefficients, because in practice they are sufficient
to describe all major peculiarities of a nonlinear oscillator
\cite{Land,Mook}.

Introducing the damping rate due to neutral gas friction,
$\gamma$, and using Eq. (\ref{1}), we can write the following
equation for the particle oscillation:
\begin{equation}\label{2}
\ddot z+2\gamma\dot z+\omega_0^2z+\alpha z^2+\beta z^3=
\frac{F(z)}{M}\cos\omega t.
\end{equation}
Here $F$ is the excitation force acting from the wire below the
particle. This force presumably depends on the vertical position
of the particle, and we expand it up to quadratic terms,
\begin{equation}\label{13}
{\textstyle F(z)\simeq F_0+ F_0^{'}z+\frac12F_0^{''}z^2}.
\end{equation}

First, let us consider the primary resonance $\omega\simeq
\omega_0$. Nonlinear effects are stronger when the oscillation
amplitude is larger, and therefore we are interested in a narrow
region $\epsilon=\omega-\omega_0$ around the resonance frequency,
where $|\epsilon|\ll\omega_0$. This approach is normally valid in
the limit $\gamma\ll\omega_0$ and is justified for our case: From
the expression for the neutral friction force \cite{Eps} we
evaluate $\gamma/2\pi\simeq0.068$~Hz which is much less than the
measured $\omega_0/2\pi\simeq17.0$~Hz. Then, we obtain from Eq.
(\ref{2}),
\begin{equation}\label{12}
\ddot z+2\gamma\dot z+\omega_0^2z+\alpha z^2+\beta z^3=\omega_0^2A
\cos(\omega_0+\epsilon)t,
\end{equation}
where $A=F_0/M\omega_0^2\equiv kU_{\rm ex}$ and $k$ is a ``scale
factor'' of the oscillation amplitude, normalized to the
excitation voltage. In Eq. (\ref{12}) we omit the terms of higher
order in $F(z)$, because for the primary resonance they are not
important when the inequality $A|F_0^{'}/F_0|\ll1$ is satisfied
(this condition always holds in our case). In accordance with the
method of successive approximations for anharmonic oscillations
\cite{Land,Mook}, the first term in the solution of Eq. (\ref{12})
is $z(t)\simeq a\cos\left(\omega_0+ \epsilon\right)t$, and the
dependence of the amplitude $a$ on $\epsilon$ and $A$
(frequency-response equation) is given by
\begin{equation}\label{3}
{\textstyle a^2\left[(\epsilon-\kappa a^2)^2+\gamma^2\right]=
\frac14\omega_0^2A^2},
\end{equation}
where the constant $\kappa$ characterizes a nonlinear shift of the
primary resonance frequency,
\begin{equation}\label{4}
\kappa=\frac{3\beta}{8\omega_0}-\frac{5\alpha^2}{12\omega_0^3}.
\end{equation}
The least-squares fit of the experimental points for the
low-amplitude (linear) primary resonance gives us the following
values: the resonance frequency $\omega_0/2\pi=17.0$~Hz, the
damping rate $\gamma/2\pi=0.067$~Hz, and the amplitude scale
factor $k=0.042$~mm/V (note that the fitted damping rate is very
close to $\gamma/2\pi=0.068$~Hz evaluated from the theory
\cite{Eps}). Using these values, we fit the points for the
nonlinear oscillations in Fig.~\ref{f2} to the function
$a(\epsilon,A)$ from Eq. (\ref{3}). The solid lines represent the
least-squares fit with $\kappa/2\pi= -0.96$~Hz/mm$^2$. If each
curve is fitted independently, this value varies within $5\%$. The
hysteresis appears on the curves, when $A$ exceeds the critical
value: $A_{\rm cr}^2 =\frac{32}{3 \sqrt{3}}(\gamma^3/|\kappa|
\omega_0^2)\simeq(2.4 \times 10^{-3}~{\rm mm})^2$, which
corresponds to $U_{\rm ex}\simeq 60$~mV.

For the superharmonic resonance $\omega\simeq\frac12\omega_0$, we
set $\epsilon=\omega-\frac12\omega_0$. The solution of Eq.
(\ref{2}) is approximately a sum of the first- and second-order
terms \cite{Land}, $z(t)\simeq z_1(t)+z_2(t)$. The first (linear)
term is $z_1(t)=a_1\cos(\frac12\omega_0+\epsilon)t$, where
\begin{equation}\label{5}
{\textstyle a_1\simeq\frac43\left(1+\frac43\epsilon/\omega_0
\right)A},
\end{equation}
represents the amplitude offset (dashed line in Fig.~\ref{f3}),
and $z_2$ describes the resonance peak due to nonlinearity.
Substituting the expression for $z_1(t)$ together with Eq.
(\ref{13}) in Eq. (\ref{2}), we get the resonance terms
$z_1^2\propto\cos(\omega_0+2\epsilon)t$ in the resulting equation
(see Ref.~\cite{Land}). Retaining these terms, we obtain,
\begin{equation}\label{6}
{\textstyle\ddot z_2+2\gamma\dot z_2+\omega_0^2z_2+\alpha z_2^2+
\beta z_2^3=-\frac89\alpha_{(\omega/2)}A^2\cos\left(\omega_0+
2\epsilon\right)t},
\end{equation}
where $\alpha_{(\omega/2)}=\alpha-\frac34\omega_0^2/\ell$ and
$\ell=F_0/F_0^{'}<0$ is the spatial scale of change of the
excitation force. In Eq. (\ref{6}) we omit the small terms $O(A^2/
\ell^2)$. Formally, the derived equation for the superharmonic
resonance is similar to that for the primary resonance [see Eq.
(\ref{12})], and therefore shapes of the corresponding curves are
qualitatively the same. Assuming $z_2(t)\simeq a_2\cos\left(
\omega_0+2\epsilon \right)t$, we obtain the following
frequency-response equation for the superharmonic resonance:
\begin{equation}\label{7}
a_2^2\left[(2\epsilon-\kappa a_2^2)^2+\gamma^2\right]=
\frac{16}{81}\left(\frac{\alpha_{(\omega/2)}}{\omega_0}
\right)^2A^4.
\end{equation}
The independent least-squares fit of the experimental points in
Fig.~\ref{f3} using Eq. (\ref{7}) gives for each curve values of
$\omega_0$ and $\gamma$, which are nearly the same as those for
the primary resonance (deviation within $\simeq0.5\%$), $\kappa/
2\pi=-1.0\pm5\%$~Hz/mm$^2$, and $|\alpha_{(\omega/2)}|/(2\pi)^2=
32\pm10\%$~Hz$^2$/mm. The critical value of $A$ for the
hysteresis: $A_{\rm cr}^4=\frac{9\sqrt{3}}{2} (\gamma^3\omega_0^2
/|\kappa|\alpha_{(\omega/2)}^2) \simeq(4.5 \times 10^{-2}~{\rm
mm})^4$, which corresponds to $U_{\rm ex}\simeq 3.5$~V.

Next, for the subharmonic resonance $\omega\simeq2\omega_0$, we
set $\epsilon=\omega-2\omega_0$. The first-order term is
$z_1(t)=-a_1\cos(2\omega_0+\epsilon)t$ with the amplitude offset
\begin{equation}\label{8}
{\textstyle a_1\simeq\frac13\left(1-\frac43\epsilon/\omega_0
\right)A}.
\end{equation}
Retaining the necessary terms (see Ref.~\cite{Land}), we derive
from Eq. (\ref{2}) the equation for $z_2$,
\begin{equation}\label{9}
{\textstyle \ddot z_2+2\gamma\dot z_2+\omega_0^2\left[1-\frac23
\left(\alpha_{(2\omega)}A/\omega_0^2\right)\cos\left(2\omega_0+
\epsilon\right)t\right]z_2+\alpha z_2^2+\beta z_2^3=0},
\end{equation}
where $\alpha_{(2\omega)}=\alpha+\frac32\omega_0^2/\ell$. We see
that Eq. (\ref{9}) is an equation of a nonlinear parametric
oscillator, and therefore the subharmonic resonance increases due
to a parametric instability \cite{Land}. Putting $z_2(t)\simeq
a_2\cos(\omega_0+\frac12\epsilon)t$, we get the frequency-response
equation,
\begin{equation}\label{10}
a_2^2\left[({\textstyle\frac12}\epsilon-\kappa a_2^2)^2+\gamma^2
\right]= \frac1{36}\left(\frac{\alpha_{(2\omega) }}{\omega_0}
\right)^2A^2a_2^2.
\end{equation}
Solution of Eq. (\ref{10}) has the following peculiarities
\cite{Land}: If $A$ is less than the threshold value $A_{\rm
th}=6\gamma\omega_0/|\alpha_{(2\omega)}|$, then Eq. (\ref{10}) has
only the zero solution $a_2=0$, and thus just the offset
$a_1(\omega)$ can be observed in experiments. For sufficiently
high $U_{\rm ex}$ ($A>A_{\rm th}$), a non-zero solution exists for
$\epsilon_{\rm b1}<\epsilon<\epsilon_{\rm b2}$, where
$\epsilon_{\rm b1,2}=\mp2\gamma\sqrt{A^2/A_{\rm th}^2-1}$. If the
frequency increases, $a_2$ jumps upward from zero at
$\epsilon_{\rm b1}$ and then continuously decreases to zero at
$\epsilon_{\rm b2}$. On the ``way back'', $a_2$ starts from zero
at $\epsilon_{\rm b2}$ and then grows monotonically. The
least-squares fit of the points in Fig.~\ref{f4} ($U_{\rm ex}=
2.0$~V) with Eq. (\ref{10}) gives us $\kappa/2\pi=-0.95$~Hz/mm$^2$
and $|\alpha_{(2\omega)}|/(2\pi)^2=320$~Hz$^2$/mm. Thus, $A_{\rm
th}\simeq 0.02$~mm, which corresponds to $U_{\rm ex}\simeq 0.5$~V.
The fit for the center $\epsilon=0$ yields $34.85$~Hz, which is
$\simeq2\%$ larger than the double value of the resonance
frequency obtained for the primary and superharmonic resonance.

Now we can finally reconstruct the sheath potential, $\phi(z)$.
Using the fitted values $|\alpha_{(\omega/2)}|=|\alpha-\frac34
\omega_0^2/\ell|\simeq(2\pi)^2\times30$~Hz$^2$/mm and
$|\alpha_{(2\omega)}|=|\alpha+\frac32\omega_0^2/\ell|\simeq
(2\pi)^2\times320 $~Hz$^2$/mm, we evaluate the first anharmonic
coefficient and the spatial scale of the excitation force:
$\alpha/(2\pi)^2\simeq-130$~Hz$^2$/mm and $\ell\simeq-2$~mm.
Substituting this value for $\alpha$ and putting $\kappa/2\pi=
-0.95$~Hz/mm in Eq. (\ref{4}), we estimate the second anharmonic
coefficient: $\beta/(2\pi)^2 \simeq20$~Hz$^2$/mm$^2$. Then, using
the series Eq. (\ref{1}) we evaluate the potential energy of a
particle in the sheath field ($z$ in mm),
\begin{equation}\label{14}
{\textstyle{\cal U}(z)\simeq M\omega_0^2\left(-0.9z+\frac12z^2-
\frac13 0.5z^3+\frac14 0.07z^4\right)\propto\phi(z)}.
\end{equation}
We see that in our case at least the first anharmonic term results
in a significant deviation of the sheath potential from a
parabolic one, and therefore the oscillations become strongly
nonlinear when the amplitude reaches a few tenths of mm, or more.
At the same time the obtained coefficients of the series converge
rapidly ($\beta/\alpha\simeq0.15$~mm$^{-1}$), so that for $|z|
\lesssim2$~mm we can limit ourselves to expansion (\ref{11}).
Thus, Eq. (\ref{14}) provides convenient analytical expression for
the electric field in rather wide vicinity of the equilibrium
particle position.

Other experiments performed with different particle sizes and
discharge conditions (Ar and Kr, $p\simeq1-7$~Pa) confirm that the
sheath potential contains considerable anharmonic terms at low
pressures. We have also investigated, how strongly the sheath
potential is perturbed by the wire below the particle. Experiments
with a wire at the same level with a particle (horizontal
separation $\simeq11$~mm) show results, which are very similar to
those reported here.

In this Letter, we have presented one example of nonlinear
particle oscillations in the sheath of a low-pressure rf discharge
and its quantitative analysis. The proposed simple method opens a
possibility to retrieve the profile of the electric field in the
whole pre-sheath and upper part of the sheath region, using
particles of different masses (and thus levitating at different
heights). Analysis for each particle allows us to reconstruct the
electric field in a range of a few mm, and therefore it is
sufficient to use just a few different particles to get rather
precise field distribution in the whole sheath range. Hence, we
believe that experiments with excitation of nonlinear oscillations
might be an effective way to study the sheath distribution.

\begin{figure}
\caption{Experimental setup.} \label{f1}
\end{figure}

\begin{figure}
\caption{Variation of the amplitude of particle oscillations close
to the primary resonance for increasing (a) and decreasing (b)
frequency of excitation, $\omega$, and for different magnitudes of
the excitation voltage $U_{\rm ex}$: 50~mV (open circles), 100~mV
(closed circles), and 200~mV (squares). Solid lines show the
least-squares fit of the points using Eq. (\ref{3}). The vertical
dotted line indicates the position of the resonance frequency,
$\omega_0$, obtained from the fit.} \label{f2}
\end{figure}

\begin{figure}
\caption{Variation of the amplitude of particle oscillations close
to the superharmonic resonance for increasing (a) and decreasing
(b) frequency of excitation, $\omega$, and for different
magnitudes of the excitation voltage $U_{\rm ex}$: 4~V (open
circles), 6~V (closed circles), and 7~V (squares). Solid lines
show the least-squares fit of the points using Eq. (\ref{7}). The
tilted dashed lines represent the amplitude offset [Eq.
(\ref{5})]. The vertical dotted line indicates the position of the
half resonance frequency, $\frac12\omega_0$, obtained from the
fit.} \label{f3}
\end{figure}

\begin{figure}
\caption{Variation of the amplitude of particle oscillations close
to the subharmonic resonance for increasing (a) and decreasing (b)
frequency of excitation, $\omega$. The excitation voltage $U_{\rm
ex}=2$~V. Solid lines show the least-squares fit of the points
using Eq. (\ref{10}). The tilted dashed lines represent the
amplitude offset [Eq. (\ref{8})]. The vertical dotted line
indicates the position of the double resonance frequency,
$2\omega_0$, obtained from the fit.} \label{f4}
\end{figure}

\end{document}